\documentclass[10pt,oneside,pdftex,dvipsnames,a4paper]{article}

\usepackage{etex}
\reserveinserts{18}

\usepackage{a4wide}
\usepackage{setspace,xspace,makeidx,fancyvrb,relsize}
\usepackage{alltt}

\usepackage{amsmath,amsthm,amssymb}
\usepackage{parallel,amsfonts,accents}

\usepackage{amsxtra,amstext,latexsym,dsfont} 

\normalfont
\usepackage{bm}

\usepackage[comma]{natbib}

\usepackage[symbol]{footmisc}
\usepackage{perpage}
\MakePerPage[2]{footnote}

\usepackage[pdftex]{graphicx}
\usepackage[dvipsnames]{xcolor}

\usepackage{pgf,tikz}
\usetikzlibrary{snakes}
\usetikzlibrary{backgrounds}
\usetikzlibrary{arrows}
\usetikzlibrary{patterns}
\usetikzlibrary{fit}
\usetikzlibrary{calc}

\usepackage{lscape} 
\usepackage{shorttoc}

\usepackage{array,longtable,booktabs,float,rotating}
\usepackage{dcolumn,ctable}

\usepackage[labelfont={bf},ruled,labelsep=period,small]{caption}

\usepackage[flushleft]{threeparttable}

\usepackage[linkcolor=black,colorlinks=true,citecolor=black,
bookmarks=true,bookmarksopen=true,bookmarksnumbered=true,bookmarksopenlevel=0,%
hyperindex=true,pdfpagemode=UseOutlines,linktocpage=true,pdfpagelayout=TwoColumnLeft,
pdftitle=Biostatistics,pdfstartview=FitH,pdffitwindow=true,urlcolor=blue]{hyperref}

\usepackage{ragged2e}

\usepackage{multirow,multicol}

\usepackage{fancyhdr,extramarks}

 \makeatletter
 \newcommand\sub{\@startsection%
     {subsubsection}{5}{0mm}{-1\baselineskip}{.01\baselineskip}%
     {\normalfont\itshape}}
 \renewcommand\subsubsection{\@startsection%
     {subsubsection}{3}{0mm}{-1\baselineskip}{.01\baselineskip}%
     {\normalfont\itshape}}
 \makeatother

\makeatletter
        \newcommand\Appendix[2][?]{%
            \refstepcounter{section}%
            \addcontentsline{toc}{appendix}%
                {\protect\numberline{\appendixname~\thesection}#1}%
            {\raggedleft\bfseries \appendixname\
                \thesection\par \centering#2\par}%
                \sectionmark{#1}%
                \@afterheading
                \addvspace{\baselineskip}}
        \newcommand\sAppendix[1]{%
            \raggedleft\bfseries\appendixname\par
            \@afterheading\addvspace{\baselineskip}}
    \makeatother

\normalsize \makeindex

\newcolumntype{A}{>{\centering}p{100pt}}
\newlength\savedwidth

\def\coldot{.}%
{\catcode`\.=\active%
    \gdef.{$\egroup\setbox2=\hbox to \dimen0 \bgroup$\coldot}}
\def\rightdots#1{%
    \setbox0=\hbox{$1$}\dimen0=#1\wd0%
    \setbox0=\hbox{$\coldot$}\advance\dimen0 \wd0%
    \setbox2=\hbox to \dimen0 {}%
    \setbox0=\hbox\bgroup\mathcode`\.="8000 $}
\def\endrightdots{$\hfil\egroup\box0\box2}

\newcolumntype{d}[1]{D{.}{.}{#1}}
\newcolumntype{A}{>{\centering}p{100pt}}
\newcolumntype{.}{D{.}{.}{-1}}
\newcolumntype{P}[2]{>{#1\raggedright\arraybackslash}p{#2}}

\DeclareFontFamily{U}{euc}{}
\DeclareFontShape{U}{euc}{m}{n}{<-6>eurm5<6-8>eurm7<8->eurm10}{}%

\theoremstyle{plain}      
\theoremstyle{plain}      
\theoremstyle{plain}      
\theoremstyle{definition} 
\theoremstyle{definition} 
\theoremstyle{definition} 
\theoremstyle{plain} 
\theoremstyle{definition} 
\theoremstyle{plain} 
\theoremstyle{definition} 

\newcounter{nctr}
\usepackage[rightbars,color]{changebar}
\cbcolor{blue}

\VerbatimFootnotes

\usepackage{url}
\newcommand\tb{\textbf}
\newcommand\ti{\textit}

\newcommand\te{\text}
\newcommand\ma[1]{\te{\bf{#1}}}
\newcommand\ca{\mathcal}

\newcommand\op{\operatorname}

\newcommand\argmax{\operatornamewithlimits{argmax}}

\newcommand\lb{\lbrace}
\newcommand\lt{\left}

\newcommand\qq{\qquad}

\newcommand\rb{\rbrace}
\newcommand\rt{\right}

\newcommand\tth{^\text{th}}

\newcommand\bk{\ma{k}} %
\newcommand\bA{\ma{A}} 
\newcommand\bD{\ma{D}} 
\newcommand\bW{\ma{W}} 

\newcommand\cE{\ca{E}} 
\newcommand\cV{\ca{V}} 
\newcommand\cW{\ca{W}} %
\usepackage{fancyhdr}
\begin{document}
\sloppy
\begin{center}
Running Head: \uppercase{Recursive Shortest Path Algorithm}
\end{center}
\vspace{3cm}

\begin{center}
\Large{\tb{Recursive Shortest Path Algorithm with \\
          Application to Density-integration of Weighted Graphs}}
\\
\vspace{2.5cm} \normalsize Cedric E. Ginestet${^{\dag\ddag}}$ and Andrew Simmons${^{\dag\ddag}}$
\end{center}
\begin{center}
  \vspace{1cm} 
  \rm ${^\dag}$ King's College London, Institute of Psychiatry, Centre for Neuroimaging Sciences (CNS) \\
  \rm $^\ddag$ National Institute of Health Research (NIHR) Biomedical
  Research Centre for Mental Health at South London and King's College
  London Institute of Psychiatry \\
\end{center}
\vspace{9cm}
Correspondence concerning this article should be sent to
Centre for Neuroimaging Sciences, NIHR Biomedical Research Centre,
Institute of Psychiatry, Box P089, King's College London, 
De Crespigny Park, London, SE5 8AF, UK. Email may be sent to \rm
cedric.ginestet@kcl.ac.uk

\pagebreak
\onehalfspacing

\begin{abstract}  
   Graph theory is increasingly commonly utilised in genetics, proteomics and
   neuroimaging. In such fields, the data of interest generally
   constitute weighted graphs. Analysis of such weighted graphs
   often require the integration of topological metrics
   with respect to the density of the graph. Here, density refers to the
   proportion of the number of edges present in that graph. When
   topological metrics based on shortest paths are of interest, such
   density-integration usually necessitates the iterative application of 
   Dijkstra's algorithm in order to compute the shortest path matrix at each
   density level. In this short note, we describe a recursive shortest path
   algorithm based on single edge updating, which replaces the need
   for the iterative use of Dijkstra's algorithm. 
   Our proposed procedure is based on pairs of breadth-first searches
   around each of the vertices incident to the edge added at each
   recursion. An algorithmic analysis of the proposed technique is
   provided. When the graph of interest is coded as an adjacency list,
   our algorithm can be shown to be more efficient than an iterative
   use of Dijkstra's algorithm. 
\end{abstract}

\section{Introduction}
The last ten years has seen a surge of interest in graph theory among
biologists, physicists and other natural scientists. This was
primarily stimulated by the seminal papers of \citet{Watts1998} and
\citet{Barabasi1999}. In particular, a wide range of different data
types are now analyzed through systematic calculations of various
topological measures, such as the characteristic path length or
clustering coefficient. In systems biology and neuroscience,
subject-specific networks can be constructed in order to compare
several populations of networks for testing putative
differences between groups of subjects \citep[see][for a
review]{Bullmore2009}. (For convenience, the terms network and graph
will here be used interchangeably, as this reflects some of the recent
developments in the literature.) Such biological networks, however, tend to be weighted
undirected graphs, which generally correspond to some standardized
covariance matrices between a set of regions of interest.
By contrast, most of the topological measures introduced by
\citet{Watts1998} and \citet{Barabasi1999} pertain to 
\ti{unweighted} networks. 

There is currently no general consensus on how to compute or compare
the topology of weighted graphs. This is a particularly arduous
problem, since it requires the use of real-valued mathematical tools
on objects, which are essentially discrete. One of the possible
solutions to this conundrum has been advanced by 
\citet{He2009a}, who suggested integrating the topological measures of
interest with respect to the \ti{density} of the network
\citep[see also][]{Achard2007,Ginestet2011a}. The density of a network
is here defined as the proportion of the number of edges in a
given graph. Such integration, however, is computationally
expensive, and its complexity grows quadratically with the number of
nodes. A Monte Carlo scheme has been proposed in the literature to 
address this issue and approximate the value of such an integral
\citep{GinestetPlos}. Such Monte Carlo methods, however, also
necessitates large number of simulations in order to reduce the
variability of the resulting estimates. 

Most of the topological metrics of interest to researchers in
neuroscience and systems biology tend to involve the computation of
the matrix of shortest paths, denoted $\bD$. This includes, for
instance, the global and local efficiency measures proposed by
\citet{Latora2001} \citep[see also][]{Latora2003}. The computation of
$\bD$ for a given network can be done
efficiently using the celebrated Dijstra's algorithm \citep{Dijkstra1959}. However,
when considering weighted networks, Dijstra's algorithm may need to be
invoked as many times as the number of edges in the graph of
interest. In this short note, we address this specific problem by
proposing a recursive shortest path algorithm based on applying
single edge updates to $\bD$. In this setup, we only work with the
shortest path matrix and compute the value of the desired topological
metric at every density level. Taken together, we therefore provide an
efficient algorithm for the density-integration of the topological
functions of weighted networks. 

\section{Density-integration of Topological Metrics}\label{sec:density-integration}
In this paper, our main focus will be on undirected weighted graphs,
containing no graph loops or multiple edges. However,
since we also need to refer to unweighted graphs, we introduce
the following notation. A graph $G$ is here defined as a triple
$(\cV,\cE,\cW)$, where $\cV(G)$ is the standard vertex set, $\cE(G)$
is the edge set and $\cW(G)$ is a \ti{multiset} of real-valued
weights. Our convention generalizes to directed graphs. In
addition, this also includes
undirected unweighted (simple) graphs as special cases, for which
the elements of $\cW$ belong to $\lb 0,1\rb$. Such
a setup may, for instance, apply to the consideration of correlation
matrices or other matrices of similarity
measures with real-valued entries. In addition, we will make use of
the following notation, 
\begin{equation}
    N_{V}:=|\cV(G)|, \qq N_{E}:=|\cE(G)|, 
    \qq\text{and}\qq N_{I}:=\frac{N_{V}(N_{V}-1)}{2},
    \notag
\end{equation}
where $:=$ signifies that the left-hand side is defined as the
right-hand side. We define $N_{I}$ as the number of shortest paths in
$G$. Naturally, $N_{I}$ here takes this value because $G$ is
undirected. For a directed network, $N_{I}$ would be $N_{V}(N_{V}-1)$.
For convenience, we will interchangeably use the following two sets of
indices to label the elements of $\cW$,
\begin{equation}
  \cW(G) = \lb w_{v_{1}v_{2}},\ldots,w_{ij},\ldots,w_{N_{V}-1, N_{V}}\rb
         = \lb w_{1},\ldots,w_{e},\ldots,w_{N_{E}}\rb.
\end{equation}
Albeit we will here restrict our attention to undirected graphs, an
extension of our proposed technique to directed networks will be
discussed in the conclusion.

A range of topological metrics necessitating the computation of the
shortest path matrix have been proposed in the literature. Two popular
choices of topological measures are the global and local efficiency
measures introduced by \citet{Latora2001}. Both of these quantities can be
derived from the general definition of the efficiency, $E(\cdot)$, of a
simple graph $G=(\cV,\cE,\cW)$, which is defined as follows,
\begin{equation}
     E(G) := \frac{1}{N_{I}}\sum_{i<j}^{N_{V}} d_{ij}^{-1},
\end{equation}
where summation over $i<j$ implies the consideration of all the
elements of the following set, $\lb i<j:
i,j=1,\ldots,N_{V}\rb$, and with $d_{ij}$ denoting the length of the
shortest path between vertices $v_{i}$ and $v_{j}$ in $G$. According
to \citet{Latora2001}, the global and local efficiencies of an
unweighted undirected graph are respectively defined as follows, 
\begin{equation}
     E^{\op{Glo}}(G):= E(G), \qq\text{and}\qq 
     E^{\op{Loc}}(G):= \frac{1}{N_{V}}\sum_{i=1}^{N_{V}}E(G_{i}), 
\end{equation}
where $G_{i}\subseteq G$ for every $i=1,\ldots,N_{V}$, such that 
each $G_{i}$ is the subgraph of all the neighbors of the $i\tth$
node. That is, $\cV(G_{i})= \lb v_{j}\in G_{i}: v_{j}v_{i}\in \cE(G)\rb$.

The computation of the efficiency or any other topological function of
$G$, which we will denote by $T(G)$ is ill-defined for a weighted
graph $G=(\cV,\cE,\cW)$. In such a case, one may resort to integrating
the topological measure of interest with respect to all the possible
densitys of the graph under scrutiny, where the density of an unweighted
undirected graph is defined as follows, 
\begin{equation}
    K(G):=\frac{N_{E}}{N_{I}}. 
\end{equation}
Now, integrating a topological function with respect to the different densitys of $G$ can
be expressed as
\begin{equation}
       \overline{T}(G):=\int_{\Omega_{K}} T(\gamma(G,k))dk,
       \label{eq:density-integrated topology}
\end{equation}
where the function $\gamma(G,k)$ in equation (\ref{eq:density-integrated topology})
is a density-thresholding function, which takes a weighted network as
well as a specific level of density and returns an unweighted network
with density $k$. Here, density is treated as a discrete random variable, $K$, with
realizations in lower case. As $K$ is discrete, it only takes a countably finite
number of values, which is the following set, 
\begin{equation}
         \Omega_{K} := \lt\lb \frac{1}{N_{I}},\frac{2}{N_{I}},
           \ldots,\frac{N_{I}-2}{N_{I}},\frac{N_{I}-1}{N_{I}},1.0\rt\rb
           = :\bk,
     \label{eq:densitys}
\end{equation}
where $|\Omega_{K}|=N_{I}$. It will also be useful to label the
elements of $\bk$ with the following indices $t=1,\ldots,N_{I}$.
Therefore, equation (\ref{eq:density-integrated topology})
can be re-written as follows, 
\begin{equation}
    \overline{T}(G) = \frac{1}{N_{I}}\sum_{t=1}^{N_{I}}
    T(\gamma(G,k_{t})). 
\end{equation}

If the topological metric of interest involves the computation of the
matrix of shortest paths, $\bD$ for every thresholding of $G$, such an
integration would necessitate invoking Dijstra's algorithm or an equivalent
method $N_{I}$ times. In the next section, we propose an alternative
to this computationally expensive integration by directly updating
$\bD$ instead of updating the underlying adjacency matrix, $\bA$, for
every new density.

\section{Recursive Shortest Path Algorithm for Density-integration}
Our strategy for bypassing the need to invoke a shortest path
algorithm at every density level consists of three stages: (i) we compute
the ranks of the entries of the weight matrix, (ii) we update the
shortest path matrix by successively adding edges in the order
corresponding to the ranks obtained in the first stage, and finally
(iii) we collect the values of the topological metric of interest for every
shortest path matrix and return the mean value of that
topological metric. We describe these three stages, in turn.

Firstly, we compute the ranks for the weighted network of interest
$G=(\cV,\cE,\cW)$ as follows, 
\begin{equation}
     R_{ij}(\cW) := \sum_{u>v}^{N_{V}} I\lb w_{ij}\geq w_{uv}\rb,
\end{equation}
where $I\lb\cdot\rb$ is the indicator function returning 1 if the
argument is true and 0 otherwise. We will assume that are no ties in
the values of $\bW$. In practice, the presence of ties can be
resolved by randomization. 

Secondly, we extract each edge in the order provided by the ranks. That
is, running over the ranks $k_{t}$, where $t=1,\ldots,N_{I}$, we have
the following $N_{I}$ ordered pairs:
\begin{equation}
      \lb v_{1},v_{2}\rb_{t} := \argmax_{\lb i,j\rb} I\lb
      P_{ij}(\bW)=k_{t}\rb.
\end{equation}
It then suffices to update $\bD$ using each of these pairs
recursively, as follows,
\begin{equation}
    \bD_{t} = \op{edgeUpdate}\lt( \bD_{t-1}, \lb v_{1},v_{2}\rb_{t}\rt).
    \label{eq:edge.update}
\end{equation}
For each $\bD_{t}$, we can now collect the topological
measure based on this particular shortest path matrix, $T(D_{t})$. 
Finally, is then remains to compute the mean value of these collected
topological measures in order to obtain the desired density-integrated
metric of the graph of interest. That is, 
\begin{equation}
     \overline{T}(G) = \frac{1}{N_{I}} \sum_{t=1}^{N_{I}} T(\bD_{t}). 
\end{equation}

\begin{figure}[t]
\tikzstyle{background rectangle}=[draw=gray!50,fill=gray!20,rounded corners=1ex]
\centering
\begin{tikzpicture}
[show background rectangle,scale=.5,every text node part/.style={font=\footnotesize},
matrix of nodes/.style={     
execute at begin cell=\node\bgroup,
execute at end cell=\egroup;,%
execute at empty cell=\node{$\infty$};,
}]
\draw[gray!20] (-18,0) -- (9,0); 
\draw[gray!20] (0,9) -- (0,-0.5); 

\draw (4.2,6.7) node[anchor=south]{$\bD$}; 
\matrix [matrix of nodes, matrix anchor=south west] at (0,-1)
{
$v_{1}$ & $v_{2}$  & $v_{3}$  & $v_{4}$  & $v_{5}$  & $v_{6}$  & $v_{7}$  \\
. & 1 & 3 & 2 &   &   &   \\
1 & . & 2 & 1 &   &   &   \\
3 & 2 & . & 1 &   &   &   \\
2 & 1 & 1 & . &   &   &   \\
  &   &   &   & . & 1 & 2 \\
  &   &   &   & 1 & . & 1 \\
  &   &   &   & 2 & 1 & . \\
};
\draw[thick,black] (0,-0.5) -- (0,5.5); 
\draw[thick,black] (8.5,-0.5)-- (8.5,5.5); 

\draw(-18,8)node[anchor=west]{\tb{a) \underline{Update:} New edge between $v_{4}$ and $v_{5}$.}}; 

    \draw[very thick,red] (-10,3)--(-7.5,3);
    \draw[thick] (-7.5,3)--(-5,1);
    \draw[thick] (-5,1) --(-2.5,3);
    \draw[thick] (-10,3)--(-12.5,1);
    \draw[thick] (-10,3)--(-12.5,5);
    \draw[thick] (-12.5,5)--(-15,3);
    \fill[black](-2.5,3)  circle(6.5pt)node[anchor=south]{$v_{7}$};
    \fill[black](-5,1) circle(6.5pt)node[anchor=south]{$v_{6}$};
    \fill[red](-7.5,3)  circle(6.5pt)node[anchor=south,black]{$v_{5}$};
    \fill[red](-10,3) circle(6.5pt)node[anchor=south,black]{$v_{4}$};
    \fill[black](-12.5,1) circle(5pt)node[anchor=south]{$v_{3}$};
    \fill[black](-12.5,5)  circle(5pt)node[anchor=south]{$v_{2}$};
    \fill[black](-15,3)  circle(5pt)node[anchor=south]{$v_{1}$};
\end{tikzpicture} \\
\begin{tikzpicture}
[show background rectangle,scale=.5,every text node part/.style={font=\footnotesize},
matrix of nodes/.style={     
execute at begin cell=\node\bgroup,
execute at end cell=\egroup;,%
execute at empty cell=\node{$\infty$};,
row 5 column 5/.style={red},row 6 column 4/.style={red},
row 5 column 6/.style={orange},row 7 column 4/.style={orange},
row 5 column 7/.style={purple},row 8 column 4/.style={purple}
}]
\draw[gray!20] (-18,0) -- (9,0); 
\draw[gray!20] (0,9) -- (0,-0.5); 

\draw (4.2,6.7) node[anchor=south]{$\bD$}; 
\matrix [matrix of nodes, matrix anchor=south west] at (0,-1)
{
$v_{1}$ & $v_{2}$  & $v_{3}$  & $v_{4}$  & $v_{5}$  & $v_{6}$  & $v_{7}$  \\
. & 1 & 3 & 2 &   &   &   \\
1 & . & 2 & 1 &   &   &   \\
3 & 2 & . & 1 &   &   &   \\
2 & 1 & 1 & . & 1 & 2 & 3 \\
  &   &   & 1 & . & 1 & 2 \\
  &   &   & 2 & 1 & . & 1 \\
  &   &   & 3 & 2 & 1 & . \\
};
\draw[thick,black] (0,-0.5) -- (0,5.5); 
\draw[thick,black] (8.5,-0.5)-- (8.5,5.5); 

\draw(-18,8)node[anchor=west]{\tb{b) \underline{Phase I:} Breadth-first search around $v_{5}$.}}; 

    \draw[very thick] (-10,3)--(-7.5,3);
    \draw[thick] (-7.5,3)--(-5,1);
    \draw[thick] (-5,1) --(-2.5,3);
    \draw[thick] (-10,3)--(-12.5,1);
    \draw[thick] (-10,3)--(-12.5,5);
    \draw[thick] (-12.5,5)--(-15,3);
    \draw[thick, red,dashed] (-10,3) to [out=-90,in=-90] (-7.5,3);
    \draw[thick, orange,dashed] (-10,3) to [out=-90,in=-135](-5,1);
    \draw[thick, purple,dashed] (-10,4) to [out=45,in=135]  (-2.5,3);
    \fill[purple](-2.5,3)  circle(6.5pt)node[anchor=south,black]{$v_{7}$};
    \fill[orange](-5,1) circle(6.5pt)node[anchor=south,black]{$v_{6}$};
    \fill[red](-7.5,3)  circle(6.5pt)node[anchor=south,black]{$v_{5}$};
    \fill[blue](-10,3) circle(6.5pt)node[anchor=south,black]{$v_{4}$};
    \fill[black](-12.5,1) circle(5pt)node[anchor=south,black]{$v_{3}$};
    \fill[black](-12.5,5)  circle(5pt)node[anchor=south,black]{$v_{2}$};
    \fill[black](-15,3)  circle(5pt)node[anchor=south,black]{$v_{1}$};
\end{tikzpicture} \\
\begin{tikzpicture}
[show background rectangle,scale=.5,every text node part/.style={font=\footnotesize},
matrix of nodes/.style={     
execute at begin cell=\node\bgroup,
execute at end cell=\egroup;,%
execute at empty cell=\node{$\infty$};,
row 6 column 1/.style={purple},row 2 column 5/.style={purple},
row 7 column 1/.style={purple},row 2 column 6/.style={purple},
row 8 column 1/.style={purple},row 2 column 7/.style={purple}, 
row 6 column 2/.style={orange},row 3 column 5/.style={orange},
row 7 column 2/.style={orange},row 3 column 6/.style={orange},
row 8 column 2/.style={orange},row 3 column 7/.style={orange},
row 6 column 3/.style={orange},row 4 column 5/.style={orange},
row 7 column 3/.style={orange},row 4 column 6/.style={orange},
row 8 column 3/.style={orange},row 4 column 7/.style={orange}
}]
\draw[gray!20] (-18,0) -- (9,0); 
\draw[gray!20] (0,9) -- (0,-0.5); 

\draw (4.2,6.7) node[anchor=south]{$\bD$}; 
\matrix [matrix of nodes, matrix anchor=south west] at (0,-1)
{
$v_{1}$ & $v_{2}$  & $v_{3}$  & $v_{4}$  & $v_{5}$  & $v_{6}$  & $v_{7}$  \\
. & 1 & 3 & 2 & 3 & 4 & 5 \\
1 & . & 2 & 1 & 2 & 3 & 4 \\
3 & 2 & . & 1 & 2 & 3 & 4 \\
2 & 1 & 1 & . & 1 & 2 & 3 \\
3 & 2 & 2 & 1 & . & 1 & 2 \\
4 & 3 & 3  & 2 & 1 & . & 1 \\
5 & 4 & 4 & 3 & 2 & 1 & . \\
};
\draw[thick,black] (0,-0.5) -- (0,5.5); 
\draw[thick,black] (8.5,-0.5)-- (8.5,5.5); 

\draw(-18,9)node[anchor=west]{\tb{c) \underline{Phase II:} Breadth-first search around $v_{4}$.}}; 

    \draw[very thick] (-10,3)--(-7.5,3);
    \draw[thick] (-7.5,3)--(-5,1);
    \draw[thick] (-5,1) --(-2.5,3);
    \draw[thick] (-10,3)--(-12.5,1);
    \draw[thick] (-10,3)--(-12.5,5);
    \draw[thick] (-12.5,5)--(-15,3);
    \draw[thick, orange,dashed] (-12.5,1) to [out=0,in=-90](-7.5,3);
    \draw[thick, orange,dashed] (-12.5,1) to [out=-15,in=-165](-5,1);
    \draw[thick, orange,dashed] (-12.5,1) to [out=-30,in=-90](-2.5,3);
    \draw[thick, orange,dashed] (-12.5,5) to [out=0,in=90](-7.5,3.5);
    \draw[thick, orange,dashed] (-12.5,5) to [out=20,in=90](-5,1.5);
    \draw[thick, orange,dashed] (-12.5,5) to [out=25,in=145](-2.5,3.5);

    \draw[thick, purple,dashed] (-15,4) to[out=90,in=180] (-11.5,7) to [out=0,in=90](-7.5,3.5);
    \draw[thick, purple,dashed] (-15,4) to[out=90,in=180] (-9.5,7.2) to [out=0,in=90](-5,1.5);
    \draw[thick, purple,dashed] (-15,4) to[out=90,in=180] (-8.5,7.5) to[out=0,in=90](-2.5,3.5);

    \fill[blue](-2.5,3)  circle(6.5pt)node[anchor=south,black]{$v_{7}$};
    \fill[blue](-5,1) circle(6.5pt)node[anchor=south,black]{$v_{6}$};
    \fill[blue](-7.5,3)  circle(6.5pt)node[anchor=south,black]{$v_{5}$};
    \fill[black](-10,3) circle(6.5pt)node[anchor=south,black]{$v_{4}$};
    \fill[orange](-12.5,1) circle(5pt)node[anchor=south,black]{$v_{3}$};
    \fill[orange](-12.5,5)  circle(5pt)node[anchor=south,black]{$v_{2}$};
    \fill[purple](-15,3)  circle(5pt)node[anchor=south,black]{$v_{1}$};
\end{tikzpicture} \\
  \caption{Graphical representation of the edge updating algorithm to
    modify the shortest path matrix, $\bD$, one edge at a time. In
    panel (a), a new edge, $v_{4}v_{5}$, is added to an existing
    graph, which is otherwise composed of two disconnected
    components. In panel (b), we conduct a BFS around $v_{5}$
    with respect to $v_{4}$, updating $\bD$ accordingly with the new
    shortest paths between $v_{4}$ and $v_{5}$ and its first and
    second degree neighbors represented in red, yellow and purple,
    respectively. In panel (c), we conduct a BFS around $v_{4}$
    with respect to the vertices, which were modified in phase I of
    edgeUpdate, denoted in blue. The first and second
    degree neighbors of $v_{4}$ are here denoted in orange and purple,
    respectively. In each panel, the
    corresponding modifications in the matrix of shortest paths are
    reported on the right-hand side. The presence of a dashed line
    between two vertices indicates that we
    test whether the inclusion of $v_{4}v_{5}$ shortens the shortest
    path between these two vertices.
    \label{fig:edge update}}
\end{figure}
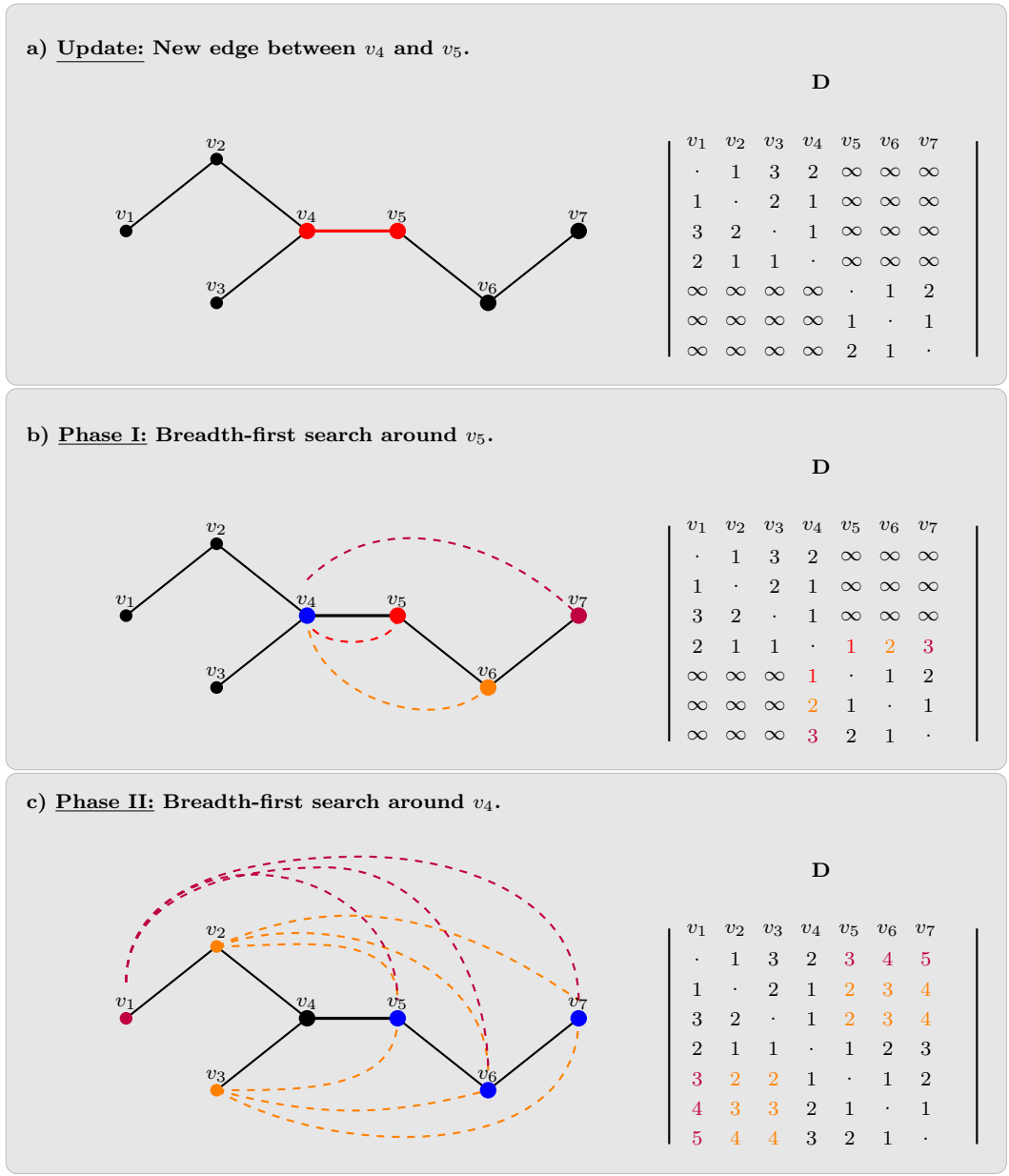

The difficulty of this method centres on the use of the edgeUpdate
function in equation (\ref{eq:edge.update}). This algorithm proceeds
as follows. At each step $t$, we ask what the impact of the
addition of a new edge to an existing graph is in terms of shortest path
relationships. Our algorithm answers this
question by two successive breadth-first searches (BFS) 
around the vertices incident to the edge added at each $t$. Firstly, we
conduct a BFS around $v_{2}$ and check whether the shortest path
between $v_{1}$ and each of the $m\tth$ degree neighbors of $v_{2}$
are shortened by the addition of a new edge between $v_{1}$ and
$v_{2}$. Secondly, we conduct a BFS centred at $v_{1}$, where we
check if the shortest paths between all the neighbors of $v_{2}$,
which were modified in the first stage and the $m\tth$ degree
neighbors of $v_{1}$ are shortened by the introduction of the new
edge. The full edge updating algorithm of $\bD_{t}$ is described in
pseudocode in Figure \ref{alg:edge update}. For simplicity, we 
represent the algorithm when each $\bD_{t}$ is coded as a full
matrix. However, a list representation can also be adopted to minimise
storage space. Moreover, we have also provided a graphical description of 
our edge updating algorithm for density-integration in Figure
\ref{fig:edge update}. A C++ version of this algorithm is freely
available as part of the NetworkAnalysis package on the R platform
(\rm http://cran.r-project.org/package=NetworkAnalysis).

\begin{figure}[t]
\small 
    \begin{Verbatim}[baselinestretch=1.15, xrightmargin=2pt,%
    frame=lines,label={[Edge Updating of $\bD$]},samepage=true,%
    commandchars=\\\{\},codes={\catcode`$=3\catcode`^=7\catcode`_=8}]

## Inputs: $\bD$, $\lb\,v_1,v_2\rb$.
## Output: $\bD$.
1    ### Initialization: 
2    Set $N_{V}=\bD\text{.ncol}()$; 
3    $d_{v_{1}v_{2}}=d_{v_{2}v_{1}}=1$; 
4
5    ### BFS around $v_{2}$:
6    Set $S_{G}=v_{1}\cup\,v_{2}$, $S^{(0)}=v_{2}$; 
7    FOR $(m=1,\ldots,N_{v}-2)$ DO
8        $\Delta=\bigcup_{v\in\,S^{(m-1)}}\,\delta(v)/S_{G}$; 
9        FOR $(v\in\,\Delta)$ DO 
10           IF $(d_{v_{1}v}\geq\,m+1)$
11              $d_{v_{1}v}=d_{vv_{1}}=m+1$; Add $v$ to $S^{(m)}$; Add $v$ to $S_{G}$; 
12           END IF; 
13       END FOR; 
14       IF $S^{(m)}=\emptyset$ BREAK; 
15   END FOR;
16
17   ### BFS around $v_{1}$:
18   Set $S_{G}=S_{G}/v_{1}$, $S^{(0)}=v_{1}$; 
19   FOR $(m=1,\ldots,N_{v}-2)$ DO 
20       $\Delta=\bigcup_{v\in\,S^{(m-1)}}\,\delta(v)/S_{G}$; 
21       FOR $(v\in\,\Delta)$ DO 
22           FOR $(u\in\,S_{G})$ DO 
23              IF $(d_{vu}\geq\,d_{v_{1}u}+m)$ 
24                 $d_{vu}=d_{uv_{1}}+m$;  $d_{vu}=d_{v_{1}u}+m$; Add $v$ to $S^{(m)}$;
25              END IF;  
26           END FOR; 
27       END FOR; 
28       IF $S^{(m)}=\emptyset$ BREAK; 
29   END FOR;
30   
31   Return \bD; 
    \end{Verbatim}
\caption{Updating of $\bD$ inserting one edge at a time, here
   denoted $v_{1}v_{2}$. The set $S_{G}$ is the set of visited
   vertices, whereas $S^{(m)}$'s are the sets of unvisited edge
   corresponding to the $m\tth$ degree neighborhoods of the previously
   modified vertices, and $\Delta$ is the set of relevant vertices at
   every level of the BFS. Both $\Delta$, $S_{G}$ and the $S^{(m)}$'s should be regarded
   as containers, where \ti{adding} implies inserting a new element in a
   set. 
   \label{alg:edge update}}
\end{figure}

\section{Algorithmic Analysis}
When storing the graph of interest as an adjacency matrix, 
Dijkstra's algorithm has efficiency in $O(|V|^{2})$. Since
density-integration would require invoking that algorithm
$N_{I}=N_{V}(N_{V}-1)/2$ times, the efficiency would, in that case, be
in $O(|V|^{4})$. If coding the graph as a matrix, our proposed algorithm
does not perform better than a combination of Dijsktra's
algorithm. As the efficiency of a BFS is $O(|V|^{2})$ and we perform
$N_{I}$ such searches, it follows that in the worst-case scenario, the
efficiency of our proposed method would also be $O(|V|^{4})$. 
However, if the graph of interest is coded as a list, each BFS is in
$O(|E|+|V|)$, and therefore the entire recursive shortest-path
algorithm has an efficiency of $O(|V|^{2}|E|+|V|^{3})$. 
By contrast, a combination of Dijstra's algorithms based on an adjacency list only
reduces to $O(|V|^{2}|E|\log|V|)$ or $O(|V|^{2}|E| + |V|^{3}\log|V|)$
using the Fibonacci heap. Thus, our algorithm outperforms a
combination of $N_{I}$ Dijkstra's algorithms when the graph of interest
is coded as a list. 


\section{Conclusion}
In this paper, we have described a recursive shortest path algorithm
for weighted graphs, which can be used for the integrating topological
metrics with respect to density. This proposed method can readily be generalized to 
directed networks. In such a case, one simply needs to define a
graph $G=(\cV,\cE,\cW)$, where the elements of $\cE(G)$ are ordered
pairs of vertices. The edgeUpdate function in equation (\ref{eq:edge.update}) can then be
modified in order to check for \ti{directed} shortest paths instead of
undirected ones. Given the growing interest of natural scientists in
graph topological properties and the large availability of weighted
networks, the utilization of algorithms of the type described in this
paper is likely to become ubiquitous.

\section{Acknowledgments} 
This work was supported by a fellowship from the UK National Institute
for Health Research (NIHR) Biomedical Research Centre for Mental
Health (BRC-MH) at the South London and Maudsley NHS Foundation Trust
and King's College London. This work has also been funded by
the Guy's and St Thomas' Charitable Foundation as well as the South
London and Maudsley Trustees. the authors also would like to thank
two reviewers for their valuable input.

\small
\singlespacing
\addcontentsline{toc}{section}{References}
\bibliography{/home/cgineste/ref/bibtex/Statistics,%
             /home/cgineste/ref/bibtex/Neuroscience}

\begin{thebibliography}{10}
\providecommand{\natexlab}[1]{#1}

\bibitem[{Achard and Bullmore(2007)}]{Achard2007}
Achard, S. and Bullmore, E. (2007).
\newblock Efficiency and cost of economical brain functional networks.
\newblock \textit{PLOS Computational Biology}, \textbf{3}, 174--182.

\bibitem[{Barabasi and Albert(1999)}]{Barabasi1999}
Barabasi, A.L. and Albert, R. (1999).
\newblock Emergence of scaling in random networks.
\newblock \textit{Science}, \textbf{286}, 509--512.

\bibitem[{Bullmore and Sporns(2009)}]{Bullmore2009}
Bullmore, E. and Sporns, O. (2009).
\newblock Complex brain networks: Graph theoretical analysis of structural and
  functional systems.
\newblock \textit{Nature Reviews Neuroscience}, \textbf{10(1)}, 1--13.

\bibitem[{Dijkstra(1959)}]{Dijkstra1959}
Dijkstra, E. (1959).
\newblock A note on two problems in connexion with graphs.
\newblock \textit{Numerische Mathematik}, \textbf{1}, 269--271.

\bibitem[{Ginestet and Simmons(2011)}]{Ginestet2011a}
Ginestet, C.E. and Simmons, A. (2011).
\newblock Statistical parametric network analysis of functional connectivity
  dynamics during a working memory task.
\newblock \textit{NeuroImage, doi:10.1016/j.neuroimage.2010.11.030},
  \textbf{5(2)}, 688--704.

\bibitem[{Ginestet et~al.(Submitted)Ginestet, Nichols, Bullmore, and
  Simmons}]{GinestetPlos}
Ginestet, C., Nichols, T., Bullmore, E., and Simmons, A. (Submitted).
\newblock Weighted network analysis: Separating differences in cost from
  differences in topology.
\newblock \textit{PLoS ONE}.

\bibitem[{He et~al.(2009)He, Dagher, Chen, Charil, Zijdenbos, Worsley, and
  Evans}]{He2009a}
He, Y., Dagher, A., Chen, Z., Charil, A., Zijdenbos, A., Worsley, K., and
  Evans, A. (2009).
\newblock Impaired small-world efficiency in structural cortical networks in
  multiple sclerosis associated with white matter lesion load.
\newblock \textit{Brain}, \textbf{132}(12), 3366--3379.

\bibitem[{Latora and Marchiori(2003)}]{Latora2003}
Latora, V. and Marchiori, M. (2003).
\newblock Economic small-world behavior in weighted networks.
\newblock \textit{The European Physical Journal B - Condensed Matter and
  Complex Systems}, \textbf{32}(2), 249--263.

\bibitem[{Latora and Marchiori(2001)}]{Latora2001}
Latora, V. and Marchiori, M. (2001).
\newblock Efficient behavior of small-world networks.
\newblock \textit{Phys. Rev. Lett.}, \textbf{87}(19), 198701--198705.

\bibitem[{Watts and Strogatz(1998)}]{Watts1998}
Watts, D.J. and Strogatz, S.H. (1998).
\newblock Collective dynamics of `small-world' networks.
\newblock \textit{Nature}, \textbf{393}(6684), 440--442.

\end{thebibliography}
\bibliographystyle{oupced3}


\addcontentsline{toc}{section}{Index}

\end{document}